\documentclass[aps,pra,twocolumn,showpacs]{revtex4}
\usepackage{graphicx,amssymb,amsthm,amsmath,bbm}

\newcommand{\be}{\begin{equation}}
\newcommand{\ee}{\end{equation}}
\newcommand{\bea}{\begin{eqnarray}}
\newcommand{\eea}{\end{eqnarray}}
\newcommand{\gammaperp}{\gamma_{\perp}}
\newcommand{\gammapar}{\gamma_{\parallel}}

\begin{document}
\title{Low-lying bifurcations in cavity quantum electrodynamics}
\author{Michael~A.~Armen}\email{armen@caltech.edu}
\author{Hideo~Mabuchi}
\affiliation{Physical Measurement and Control 266-33, California
Institute of Technology, Pasadena, California 91125 USA}

\date{\today}
\begin{abstract}
The interplay of quantum fluctuations with nonlinear dynamics is a central
topic in the study of open quantum systems, connected to fundamental issues
(such as decoherence and the quantum-classical transition) and practical
applications (such as coherent information processing and the development of
mesoscopic sensors/amplifiers). With this context in mind, we here present a
computational study of some elementary bifurcations that occur in a driven and
damped cavity quantum electrodynamics (cavity QED) model at low intracavity
photon number. In particular, we utilize the single-atom cavity QED Master
Equation and associated Stochastic Schr\"odinger Equations to characterize the
equilibrium distribution and dynamical behavior of the quantized intracavity
optical field in parameter regimes near points in the semiclassical
(mean-field, Maxwell-Bloch) bifurcation set. Our numerical results show that
the semiclassical limit sets are qualitatively preserved in the quantum
stationary states, although quantum fluctuations apparently induce phase
diffusion within periodic orbits and stochastic transitions between attractors.
We restrict our attention to an experimentally realistic parameter regime.
\end{abstract}  \maketitle

\section{Introduction}
%

\noindent Bifurcation analysis is a fundamental aspect of dynamical systems
theory \cite{Stro01,Perk06}. It provides a powerful set of tools and concepts
for the study of bistability, hysteresis and related phenomena in natural and
engineered systems. In some practical applications the theory can be used to
ensure operation in a structurally stable parameter range, while in others it
is used to identify operating points that are highly sensitive to specific
perturbations. The latter objective can arise in scenarios where, for example,
one seeks to exploit the intrinsic nonlinear dynamics of a sensing device to
provide amplification. This remarkably robust strategy applies even in
nanomechanical \cite{Almo05} and single trapped-ion \cite{Tsen99} systems, as
well as in superconducting circuit implementations of quantum computation
\cite{Sidd04}. However, the role of quantum fluctuations in determining minimum
noise figures and back-action for bifurcation amplifiers is not yet understood.
It has also long been appreciated that bistability and hysteresis could provide
a basis for designing logic devices for switching and computation in nonlinear
optics \cite{Abra82}; interest in this subject has been reinvigorated by
advances in the fabrication of photonic bandgap structures and other integrated
optical circuits \cite{Yani03,Prie05}. Implementations based on strong coupling
to quantum dots may provide access to a technologically important regime of
attojoule and picosecond switching, but such performance would seem to imply an
energy separation between logical states on the order of tens of photons. Hence
the role of quantum fluctuations again calls for attention. Bifurcation
analysis and design thus have many roles to play in modern engineering and
applied science, and there is a real need to incorporate stochastic analysis.

In recent years, there has been growing interest in studying bifurcation-like
behavior of physical, chemical and biological systems that are fundamentally
discrete and stochastic in nature but that can be well described in some
relevant limit by nonlinear differential equations or maps. In chemical
reactions, for example, ordinary differential rate equations (corresponding to
the so-called law of mass action) provide an accurate model in the limit where
the number of molecules per species becomes infinite (at fixed concentration)
\cite{Gill77}. As a cell biologist or as an engineer designing reaction
networks for molecular computation, however, one might well be concerned with
understanding precisely how bifurcation phenomena predicted by the rate
equations are reflected in the stochastic dynamics of a small number of
reacting molecules. Stochastic extensions of rigorous bifurcation theory are
now being developed by several authors \cite{Arno98,Zeem88}, but experimental
and numerical investigations of specific systems are providing crucial guidance
for early development of the field \cite{Elow00,Qian02,Steu04}.

Stochastic behavior and dynamic nonlinearity naturally coexist in the quantized
setting of cavity quantum electrodynamics (cavity QED) with strong coupling
\cite{RempeKimble91,Mabu02}. While the latter subject may seem a bit esoteric
in comparison to chemical reaction networks, it has important connections to
nonlinear-optical signal processing and offers the additional interest of
incorporating quantum interference and atom-field entanglement. As in the case
of chemical reactions, the microscopic equations of cavity QED are known to
have a continuous and deterministic macroscopic limit (corresponding to many
weakly-coupled atoms), which is related to elementary models of the laser. It
was recognized even in the early days of cavity QED \cite{Sava88} that progress
in the laboratory would eventually provide a means of exploring bifurcation
phenomena such as optical bistability in a single-atom and few-photon regime;
such experiments are indeed now feasible. An opportunity thus arises for the
empirical study of bifurcation phenomena in the discrete physical limit,
providing an intriguing quantum-optical counterpart to the systems mentioned
above.

While previous theoretical \cite{Sava88,Kili91,Kozl99} and experimental
\cite{Hood98} investigations of single-atom bistability have largely focused on
steady-state observables of the transmitted optical field, we will here follow
the spirit of Refs.~\cite{Alsi91,Well01} in studying transient signals and
stochastic jumps observable in the broadband photocurrent generated in
individual experimental trials. We look in particular at a case of absorptive
bistability, a supercritical Hopf bifurcation, and a subcritical Hopf
bifurcation, all of which occur with mean intracavity photon numbers of order
ten. Our principal aims in this paper are to illustrate a systematic approach
(building upon Ref.~\cite{GangNingHaken90,GNH90b}) to expanding the known
inventory of bifurcation-type phenomena in single-atom cavity QED, and to
highlight some conspicuous predictions of the fully quantum model as compared
to the semiclassical Maxwell-Bloch Equations. In so doing we hope to begin to
illumine a more comprehensive picture of the quantum-classical transition in
cavity nonlinear optics \cite{Berm94,Mand97}, bridging what is generally known
about linear-gaussian \cite{Wise05} and chaotic \cite{Habi06} open quantum
systems \cite{Vard01}.

\section{The driven and damped Jaynes-Cummings model}\label{sec:JCModel}
\subsection{Quantum Dynamical Description}\label{subsec:Quantum}
We consider the driven Jaynes-Cummings Hamiltonian 
\cite{CarmichaelOpenSys} which models the interaction of a single mode of an
optical cavity having resonant frequency $\omega_c$,  with a two-level atom,
comprised of a ground state $|g\rangle$ and an excited state $|e\rangle$
separated by a frequency $\omega_a$. For an atom-field coupling constant $g_0$
and a drive field amplitude $\mathcal{E}$, the Hamiltonian written in a frame
rotating at the drive frequency $\omega_l$ is given by [$\hbar = 1$]: \be
\label{eq:JCHamiltonian} \mathcal{H}=\Delta_c a^{\dag}a+\Delta_a \sigma_{+}
\sigma_{-} +i g_0 (a^{\dag}\sigma_{-} -a\sigma_{+} )+i\mathcal{E}(a^\dag-a) \ ,
\ee where $\, \Delta_a = \omega_a-\omega_l\,$ and $\, \Delta_c =
\omega_c-\omega_l\, $. In equation (\ref{eq:JCHamiltonian}), $a$ is the field
annihilation operator and $\sigma_{-}  = |g\rangle\langle e |$ is the atomic
lowering operator. In addition to the coherent dynamics governed by
(\ref{eq:JCHamiltonian}) there are two dissipative channels for the system: the
atom may spontaneously emit into modes other than the preferred cavity mode, at
a rate $\gammapar$, and photons may pass through the cavity output coupling
mirror, at a rate $2\kappa$. Furthermore, we model the case of non-radiative
dephasing (at rate $\gamma_{nr}$) between the atomic ground and excited states.
In the analysis to follow, we will be concentrating solely on the situation
where  $\gamma_{nr} = 0$, \textit{i.e.} purely radiative damping; however,
$\gamma_{nr}$ is included here to indicate that we are not restricted to this
case (in particular, the parameterization employed in section
\ref{sec:BifurcationAnalysis} will imply a variable dephasing.) The
unconditional master equation describing this driven, damped, and dephased
evolution is
\begin{eqnarray}
\dot{\rho}&=&-i [\mathcal{H},\rho]+ \kappa(2 a \rho a^\dag-a^\dag a
\rho -\rho a^\dag a ) \nonumber \\
&& + \ \gammapar/2(2 \sigma_{-}  \rho
\sigma_{+} -\sigma_{+}  \sigma_{-}  \rho -\rho \sigma_{+}  \sigma_{-} ) \label{eq:JCMaster} \\
&& + \ \gamma_{nr}/2(\sigma_z \rho \sigma_z - \rho ) \nonumber
\end{eqnarray}
where $\sigma_z = [ \sigma_{+} ,\sigma_{-} ]$ measures the population
difference between the excited and ground states.
\par
While $g_0$ measures the coherent coupling rate between the atom and the
cavity, the rates $\gammapar$, $\gamma_{nr}$, and $\kappa$ characterize
processes which tend to inhibit the build up of coherence. The qualitative
nature of the dynamics (\ref{eq:JCMaster}) may be determined by two
dimensionless parameters which measure the relative strengths of the coherent
and incoherent processes: the critical photon number
\be%
n_{0} = \frac{\gammapar\gammaperp}{4 g_0^2} \;,
\ee%
and the critical atom number
\be%
N_0 = \frac{2\gammaperp\kappa}{g_0^2} \;,
\ee%
where $\gammaperp$ is the transverse relaxation rate given by $\gammaperp =
\gammapar/2 + \gamma_{nr}$. The critical photon number provides a measure of
the number of photons needed to saturate the response of a single atom.
Therefore, in the regime $n_0 < 1$, a single photon inside the resonator may
induce a nonlinear system response. Similarly, the critical atom number roughly
quantifies the number of atoms required to drastically change the resonant
properties of the cavity. When $N_0 < 1$, a single atom inserted into the
cavity will have a dramatic effect on the cavity output. The so-called ``strong
coupling regime'' of cavity QED, which is usually used to denote the regime
where the coherent coupling dominates over dissipation, is reached when the
condition $(n_0, N_0) < 1$ holds.
\par
The master equation (\ref{eq:JCMaster}) may be used to find the time evolution
for any operator acting on the system Hilbert space. In particular, it will be
useful to know the dynamical equations for $\langle a\rangle$, $\langle
\sigma_{-}  \rangle$, and $\langle \sigma_z \rangle$ in order to make concrete
comparisons with the semi-classical results that follow. Using the fact that
$\dot{\langle O\rangle} = \mathrm{Tr}[O\dot{\rho}]$ for a system operator $O$,
we obtain
\bea%
\dot{\langle a\rangle}&=&-\kappa(1+i\Theta)\langle
a\rangle+g_0\langle\sigma_{-}\rangle+\mathcal{E}\nonumber \\
\label{eq:MBQuant}
\dot{\langle\sigma_{-} \rangle}&=& -\gamma_\perp(1+i\Delta) \langle\sigma_{-}
\rangle+g_0 \langle a \sigma_z\rangle\\
\dot{\langle\sigma_z\rangle}&=&
-\gamma_\parallel(\langle\sigma_z\rangle+1)-2g_0 (\langle
a^\dag \sigma_{-} \rangle+\langle\sigma_{+}  a\rangle) \;, \nonumber
\eea%
with $\gammaperp  = \gammapar /2 +\gamma_{nr}$, $\Theta =
(\omega_c-\omega_l)/\kappa$, and $\Delta = (\omega_a-\omega_l)/\gammaperp$.
\par
It should be noted that these formulae may be easily generalized to the case of
$N$ non-interacting atoms each coupled to the same mode of the electromagnetic
field, with coupling constant $g_0$. In this case, the Hamiltonian becomes%
\be
\label{eq:JCHamN} \mathcal{H}= \Delta_c a^{\dag}a+\sum_{j=1}^{N}
\Delta_a\sigma_{+}^j\sigma_{-}^j+\sum_{j=1}^{N}ig_0
(a^{\dag}\sigma_{-}^j-a\sigma_{+}^j )+i\mathcal{E}(a^\dag-a) \ ,
\ee%
and the new master equation is
\begin{eqnarray}
\dot{\rho}&=&-i [\mathcal{H},\rho]+ \kappa(2 a \rho a^\dag-a^\dag a
\rho -\rho a^\dag a ) \nonumber \\
&& + \ \gammapar/2\sum_{j=1}^N (2 \sigma_{-}^j \rho
\sigma_{+}^j-\sigma_{+}^j \sigma_{-}^j \rho - \rho\sigma_{+}^j \sigma_{-}^j )
\label{eq:JCMasterN}\\
&& + \ \gamma_{nr}/2\sum_{j=1}^N ( \sigma^j_z \rho \sigma^j_z - \rho )\nonumber \ ,
\end{eqnarray}%
where $\sigma_{-}^j$ is the lowering operator for the $j$th atom and $[\sigma_{+}^j,
\sigma_{-}^k] = \delta_{jk} \sigma_z^j$. The equations of motion for the operator
expectations become%
\bea
\dot{\langle a\rangle}&=&-\kappa(1+i\Theta)\langle a\rangle+g_0\sum_{j=1}^N\langle
\sigma_{-}^j \rangle+\mathcal{E}\nonumber \\ \label{eq:MBQuantNj}
\dot{\langle\sigma_{-}^j \rangle}&=& -\gamma_\perp(1+i\Delta) \langle\sigma_{-}^j
\rangle+g_0 \langle a \sigma_z^j\rangle\\
\dot{\langle\sigma_z^j\rangle}&=&
-\gamma_\parallel(\langle\sigma_z^j\rangle+1)-2g_0 (\langle
a^\dag \sigma_{-}^j \rangle+\langle\sigma_{+}^j  a\rangle) \; . \nonumber
\eea%
If we define $\ \sigma_{-} = \sum_{j=1}^{N} \sigma_{-}^j \ $ and $\ \sigma_z =
\sum_{j=1}^{N} \sigma_z^j \ $ as the collective pseudo-spin operators, we arrive
at the following set of dynamical equations%
\bea
\dot{\langle a\rangle}&=&-\kappa(1+i\Theta)\langle a\rangle + g_0\langle\sigma_{-}
\rangle+\mathcal{E}\nonumber \\ \label{eq:MBQuantN}
\dot{\langle\sigma_{-} \rangle}&=& -\gamma_\perp(1+i\Delta) \langle\sigma_{-}
\rangle+g_0 \langle a \sigma_z\rangle\\
\dot{\langle\sigma_z\rangle}&=&
-\gamma_\parallel(\langle\sigma_z\rangle+N)-2g_0 (\langle
a^\dag \sigma_{-} \rangle+\langle\sigma_{+} a\rangle) \;. \nonumber
\eea%
Therefore, we may think of equation (\ref{eq:MBQuantN}) as a description of the
operator expectation dynamics for either a single-atom (with $N=1$) or
multi-atom system. In either case, the coupled equations (\ref{eq:MBQuantN})
are not closed, as they contain expectation values of operator products.
Therefore, we also need the dynamical equations for the higher order moments,
of which there are an infinite number. For purely optical systems, order
parameters can often be identified so that a system size expansion can yield a
finite, closed set of equations which are valid in the ``low-noise" limit (when
the order parameter is large). Unfortunately, for coupled atom-field systems,
there exists no suitable choice of system scaling parameters which would
justify a system size expansion \cite{CarmichaelOpenSys}. Furthermore, it is
known that the quantum fluctuations produced by optical bistability can be
non-classical even when $N\gg1$ \cite{RempeKimble91}, and therefore would not
fit into the classical mold which is the basis of a system size expansion.
Nevertheless it has been demonstrated \cite{Rose91} that the Maxwell-Bloch
equations, which will be derived from (\ref{eq:MBQuantN}) below, can be brought
with some refinements into close agreement with experiments on absorptive
optical bistability in a multi-atom system. Indeed said equations are generally
accepted as a canonical, though somewhat phenomenological, model for cavity
nonlinear optics outside the strong coupling regime \cite{Mand97,Lugiato}.

\subsection{Semi-Classical Description}\label{subsec:Semiclassical}
An {\it ad hoc} (and somewhat crude) approach obtaining a closed set of
equations from (\ref{eq:MBQuantN}) is to simply factorize the operator
products, \textit{e.g.} $\langle a^\dag\sigma_{-} \rangle\rightarrow \langle
a^\dag\rangle\langle\sigma_{-} \rangle$. While there is no formal basis for
this procedure in general, the intuition behind it is that for a large system
with many weakly-excited atoms, the atom-field correlations will tend to zero,
allowing for expectations of operator products to be factorized
\cite{Lugiato,Carm86}. But it should be noted that this approximation is not
justified in the case of strong driving and certainly not for a single atom.
This factorization yields
\bea%
\dot{\langle{a}\rangle}&=&-\kappa\left(1+i\Theta\right)\langle a\rangle+g_0\langle
\sigma_{-} \rangle+\mathcal{E}\nonumber \\
\dot{\langle\sigma_{-} \rangle}&=& -\gamma_\perp\left(1+i\Delta\right) \langle
\sigma_{-} \rangle+g_0 \langle a \rangle\langle\sigma_z\rangle \label{eq:MBQuantFactorized}\\
\dot{\langle \sigma_z \rangle}&=&
-\gamma_\parallel\left(\langle\sigma_z\rangle+N\right)-2g_0 (\langle
a^\dag\rangle\langle\sigma_{-} \rangle+\langle\sigma_{+} \rangle
\langle a\rangle)\nonumber
\eea%
which are the well known the Maxwell-Bloch equations, used to describe the
semi-classical evolution of a classical field coupled to an atomic medium. The
atom-field correlations which were discarded in performing the factorization
above will tend to contribute ``noise'' on top of the mean field evolution
described by Eq.~(\ref{eq:MBQuantFactorized}). To put these equation into a
more common form, we make the following definitions:
\begin{eqnarray}
\tilde{x} \doteq \langle a\rangle ,\quad \tilde{p} \doteq \frac{2}{N}
\langle\sigma_{-}\rangle ,\quad \tilde{D} \doteq
\frac{1}{N}\langle\sigma_z\rangle.
\end{eqnarray}
so that (\ref{eq:MBQuantFactorized}) becomes
\begin{eqnarray}\label{eq:MBdims}
\dot{\tilde{x}} &=& -\kappa(1+i\Theta)\tilde{x}+(N g_0/2) \tilde{p}+\mathcal{E}
\nonumber \\
\dot{\tilde{p}} &=& -\gamma_\perp(1+i\Delta) \tilde{p} + 2 g_0 \tilde{x} \tilde{D}\\
\dot{\tilde{D}} &=& -\gamma_\parallel(\tilde{D}+1)-g_0(\tilde{x}^*\tilde{p}+\tilde{p}^*
\tilde{x}) \ .\nonumber
\end{eqnarray}
A computationally more practical form of (\ref{eq:MBdims}), which will prove
useful in the bifurcation analysis to follow, may be obtained by transforming
it into a dimensionless set of equations. We first make the following change of
variables
\begin{eqnarray}\label{eq:FieldScaling}
 \tilde{x} \rightarrow \sqrt{n_{0}} \, x , \quad \tilde{p} \rightarrow -
 \sqrt{\frac{\gamma_\parallel}
{\gamma_\perp}} \, p, \quad \tilde{D} \rightarrow -D, \quad
\end{eqnarray}
followed be a re-scaling of time%
\bea
t \rightarrow t' / \gamma_\perp \ ,
\eea
so that we are left with the dimensionless Maxwell-Bloch equations \cite{Lugiato}:
\begin{eqnarray}\label{eq:MBdimless}
\dot{x} &=& -k \left[(1+i\Theta) x  + 2 C p - y\right]\nonumber \\
\dot{p} &=& - (1+i\Delta) p + x D \\
\dot{D} &=& -\gamma \left[D -1 + (x^*p+p^*x)/2 \right]\nonumber
\end{eqnarray}
where the complex variables $x$ and $p$ represent the amplitude of the
intra-cavity field and the normalized atomic polarization, respectively, $D$ is
the (real) atomic population inversion, and $y$ is the amplitude of the
external drive field. The cooperativity parameter, $C$, measures the strength
of the collective atom-field interaction, while $k$ and $\gamma$ are,
respectively, the cavity field decay and atomic spontaneous emission rates,
scaled by the atomic transverse relaxation rate, $\gamma_\perp$:
\begin{eqnarray}
\gamma=\frac{\gamma_\parallel}{\gamma_\perp}, \quad
k=\frac{\kappa}{\gamma_\perp},\quad C =
\frac{N g_0^2}{2\kappa\gamma_\perp},\quad y =
\frac{\mathcal{E}}{\kappa\sqrt{n_{0}}} \ .
\end{eqnarray}
The two detuning parameters $\Theta$ and $\Delta$ are the same as in Eq.~(\ref{eq:MBQuant}).
Although there is no way to express the steady-state
solutions for the dependent variables $x$, $p$, and $D$ in terms of the independent
variables, we can find a simple set of equations relating the stationary solutions
of the problem:%
\begin{eqnarray}\label{eq:MBEOS}
y &=& |x_{ss}|\left\{\left[1+\frac{2C}{1+\Delta^2+|x_{ss}|^2}\right]^2 \right. \nonumber \\
&& + \ \left. \left[\Theta-\frac{2C\Delta}{1+\Delta^2+|x_{ss}|^2}\right]^2\right\}^{1/2} \ ,\\
p_{ss} &=& \frac{(1-i\Delta)x_{ss}}{1+\Delta^2+|x_{ss}|^2} \ , \\
D_{ss} &=& \frac{1+\Delta^2}{1+\Delta^2+|x_{ss}|^2} \ .
\end{eqnarray}%
\par
\begin{figure}
\centering 
{\includegraphics[width=0.475\textwidth]{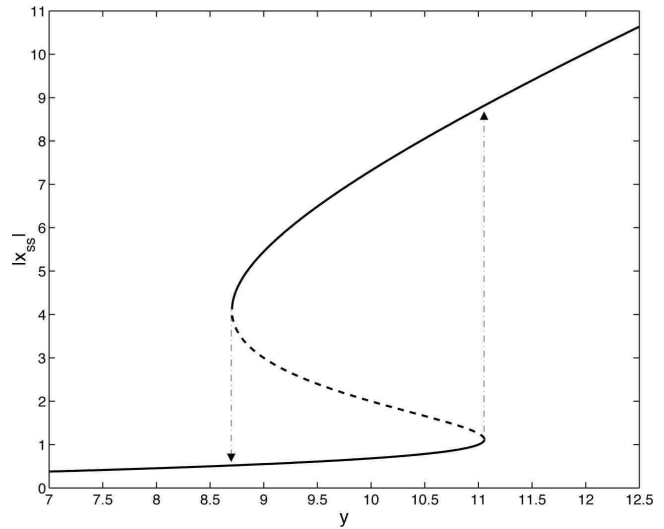}}
\caption{\label{fig:AbsorptiveBistabilityMBSim} Semiclassical calculation of
the intracavity steady-state field magnitude $|x_{ss}|$ versus drive field $y$.
The dashed portion of the curve is unstable. The parameter values are:
$C=10,k=0.1,\gamma=2,\Theta=0,$ and $\Delta=0.$ Arrows indicate the evolution
of the steady-state solution when the drive field $y$ is swept smoothly through
a bifurcation point: the state originally on the lower (upper) branch moving
through the bifurcation point $y\approx11.1$ ($y\approx8.7$) is attracted to
the upper (lower) branch. }
\end{figure}

In Fig.~\ref{fig:AbsorptiveBistabilityMBSim} we plot a typical input-output
curve generated using Eq.~(\ref{eq:MBEOS}). Note that in the range $8.7
\lesssim y\lesssim11.1$ the curve displays absorptive bistability, with the
lower and upper branches of the `S'-shaped curve supporting stable solutions
(the dashed portion of the curve is unstable).

It is important to note that the above equations depend upon $g_0$ and $N$ only
through the cooperativity parameter, $C$. Thus, identical behavior is predicted
for a range of systems with varying atom number and
$g_0=\sqrt{2\kappa\gamma_\perp C/N}$. Of course, one expects that the quantum
fluctuations and atom-field correlations that are disregarded in the derivation
of the Maxwell-Bloch equations should begin to matter as $N$ approaches 1. A
direct comparison of `system behavior' according to (\ref{eq:MBdimless}) versus
the master equation (\ref{eq:JCMaster}) with the quantities
$\{\gamma,k,C,y,\Theta,\Delta\}$ held fixed can thus be construed as a case
study in quantum-(semi)classical correspondence. The question of course is
exactly what `system behavior' should be compared and how; the strategy in what
follows will be to focus on photocurrent properties near bifurcation points of
the semiclassical model. We thus next discuss a systematic approach to finding
interesting points in the bifurcation set of the Maxwell-Bloch equations, and
then review a standard monte carlo approach to simulating photocurrents
according to the quantum model. After presenting some numerical results, we
conclude with a discussion of some interesting features of the
quantum-semiclassical comparison that suggest directions for further research.

\section{Bifurcation set of the mean-field equations}\label{sec:BifurcationAnalysis}
In this section we delineate the process used to find and classify
bifurcations in the mean-field dynamics described by Eq.~(\ref{eq:MBdimless}).
In particular, in Sec.~\ref{subsec:Linearize} we characterize both saddle-node
bifurcations and Hopf bifurcations. We further differentiate between super-
and subcritical Hopf bifurcations in Sec.~\ref{subsec:SuperSubHopf}. In the
former case, the bifurcation will destabilize a (typically) fixed point with a
\textit{local} (small amplitude) limit cycle born about the prior steady-state.
In the latter case, \textit{no local} limit cycle is created about the
destabilized steady-state solution, and the system will move to a new (possibly
distant) attractor. For this reason, subcritical Hopf bifurcations often lead
to qualitatively more radical results, including regions of multi-stability.

\subsection{Linearization about steady-state}\label{subsec:Linearize}
In order to determine the parameter values that lead to bifurcations, we return
to equation (\ref{eq:MBdimless}), and linearize the system dynamics about
steady-state. We consider small fluctuations $\delta x, \delta p$, and $\delta
D$ about steady-state, and set $x = x_{ss} +\delta x$, $p = p_{ss} +\delta p$,
{\it et cetera}. After eliminating terms that are second order in the small
fluctuations, 
\begin{equation}
\left(
\begin{array}{c}
\dot{\delta x}\\
\dot{\delta x^*}\\
\dot{\delta p}\\
\dot{\delta p^*}\\
\dot{\delta D}
\end{array}
\right) = \mathbf{J}
\left(
\begin{array}{c}
\delta x\\
\delta x^*\\
\delta p\\
\delta p^*\\
\delta D
\end{array}
\right)
\end{equation}
where the Jacobian $\mathbf{J}$ is given by
\begin{equation}\label{eq:MBJ}
\begin{split}
\mathbf{J} &= \\
&-\left(
\begin{array}{ccccc}
k(1+i\Theta)  & 0  &2Ck&0&0   \\
 0&k(1-i\Theta) &0&2Ck&0   \\
 -D_{ss}&0 &1+i\Delta&0&-x_{ss}\\
 0&-D_{ss}&0&1-i\Delta&-x_{ss}^*\\
 \gamma p_{ss}^*/2&\gamma p_{ss}/2&\gamma x_{ss}^*/2&\gamma x_{ss}/2&\gamma
\end{array}
\right).
\end{split}
\end{equation}
The associated characteristic equation will have the form
\begin{equation}\label{eq:characteristic}
\lambda^5+a_1\lambda^4+a_2\lambda^3+a_3\lambda^2+a_4\lambda+a_5 = 0
\end{equation}
with coefficients given by the following expressions:
\begin{equation}
\begin{array}{lclclc}\label{eq:CharacteristicCoeffs}
a_1 &=& 2+\gamma+2k \ ,\\
a_2 &=& k^2(1+\Theta^2)+(2\gamma+1+\Delta^2+\gamma |x_{ss}|^2) \\
& & + \ 2k(\gamma+2)+4kCD_{ss} \ ,\\
a_3 &=& \gamma(1+\Delta^2+|x_{ss}|^2)+2k(2\gamma+1+\Delta^2+\gamma |x_{ss}|^2) \\
& & + \ k^2(1+\Theta^2)(\gamma+2) + 4kCD_{ss}(\gamma+k+1)\\
& & - \ \gamma k C(p_{ss}^* x_{ss}+p_{ss} x_{ss}^*) \ ,\\
a_4 &=& 2k\gamma(1+\Delta^2+|x_{ss}|^2)\\
& & + \ k^2(1+\Theta^2)(2\gamma+1+\Delta^2+\gamma |x_{ss}|^2)\\
& & + \ 2kCD_{ss}[2k(1-\Delta\Theta)+2\gamma(k+1)+\gamma |x_{ss}|^2]\\
& & + \ \gamma k C\left[i(\Delta+k\Theta)(p_{ss}^*x_{ss}-p_{ss}x_{ss}^*)\right.\\
& & -\ \left.(k+1)(p_{ss}^* x_{ss}+p_{ss}x_{ss}^*)\right] +  4k^2C^2D_{ss}^2 \ , \\
a_5 &=& \gamma k^2\{4C^2D_{ss}[D_{ss}-(p_{ss}^* x_{ss}+p_{ss} x_{ss}^*)2]\\
& & + \ (1+\Theta^2)(1+\Delta^2+|x_{ss}|^2)\\
& & + \ 4CD_{ss}(1-\Delta\Theta)\} \ .
\end{array}
\end{equation}
These coefficients may be further simplified by using the form of $D_{ss}$
found in (\ref{eq:MBEOS}) and the relations
\begin{eqnarray*}
i(p_{ss}x_{ss}^*-p_{ss}^*x_{ss}) &=& -2\Delta |x_{ss}|^2/(1+\Delta^2+|x_{ss}|^2)\\
(p_{ss}x_{ss}^*+p_{ss}^*x_{ss}) &=& 2 |x_{ss}|^2/(1+\Delta^2+|x_{ss}|^2)\quad,
\end{eqnarray*}
so that the $a$'s found in the characteristic equation are written explicitly
in terms of six parameters: $C,k,\gamma,\Theta,\Delta,$ and $|x_{ss}|$.
\par
Needless to say, it is impossible to solve for the eigenvalues of this system
analytically. However, other methods that provide analytic tests of stability
do exist. Most notably, the Routh-Hurwitz criterion provides a set of
inequalities based on combinations of the $a's$ that can be used to determine
stability. Unfortunately, this procedure is simply too general, and it is ill
suited for the purpose of determining the boundaries of instability in terms of
our controllable parameters. We can, however, make use of the Routh-Hurwitz
criteria to find the following \textit{necessary} conditions for stability
\cite{Orozco89, HuYang88}
\[
a_1,\ a_2,\ a_3,\ a_4,\ a_5\ >\ 0 \ .
\]
Furthermore, at a Hopf bifurcation the system must have a pair of pure
imaginary eigenvalues, $\lambda_{1,2} = \pm i \omega$.  Demanding that the
characteristic equation support these solutions establishes the following
critical condition for a Hopf bifurcation
\begin{equation}\label{eq:f}
f = (a_1a_2-a_3)(a_3a_4-a_2a_5)-(a_1a_4-a_5)^2 = 0 \quad,
\end{equation}
with $f>0$ providing another necessary condition for stability.
\par
For the purpose of delineating the instability boundaries, the six inequalities
\begin{equation}\label{eq:InstabilityBoundary}
a_1,\ a_2,\ a_3,\ a_4,\ a_5,\ f\ >\ 0
\end{equation}
are not equally important. Starting from a stable region of the parameter
space, there are only two ways for the steady-state solution to become
unstable: (i) a single real eigenvalue passes through the origin and becomes
positive; (ii) a pair of complex conjugate eigenvalues cross the imaginary axis
(starting from the LHP). For case (i) the coefficient $a_5$ must change signs
first, whereas for case (ii) it is $f$ that first changes sign. Therefore, if
the goal is to determine the conditions for a known \textit{stable} state to
become unstable, there is no need to consider the other necessary conditions
and all focus may be placed on $a_5$ and $f$. Furthermore, if we are only
interested in Hopf Bifurcations, we can also ignore the $a_5>0$ condition,
which determines the boundary for saddle-node bifurcations where the
steady-state curve displays a turning point (this can be seen by noting that
$d(y^2)/d(|x_{ss}|^2) \propto a_5$, so that $a_5<0$ indicates bistability.)
\par
It should be noted again, that the inequalities (\ref{eq:InstabilityBoundary})
are only necessary conditions for stability, they are not sufficient. For
example, the system could have one real negative eigenvalue, and two pairs of
complex eigenvalues each with positive real parts, and still have
$a_1,a_2,a_3,a_4,a_5,f>0$. However, the stability condition $f>0$ can be made
sufficient for a given region in parameter space if we can show that  this
region is `connected' to a known stable region of the space (two regions of
parameter space are `connected' if there exists a continuous variation of the
parameters that moves the system from one region, while retaining the sign of
$a_1,a_2,a_3,a_4,a_5,$ and $f$ through the entire path.) Thus, if we know that
a particular region of parameter space (with $a_1,a_2,a_3,a_4,a_5>0$) is
connected to a stable region, we know that $f$ can serve as a necessary and
sufficient condition for the steady-state solution to undergo a Hopf
bifurcation. Practically, all this means is that starting from a stable state,
the first crossing of a surface $a_5=0$ ($f=0$), will drive the system unstable
through a saddle-node (Hopf) bifurcation. Furthermore, the stability condition
$f>0$ is quite reliable in practice, even when we can't show connectedness to a
stable region.

\subsection{Super- and subcritical Hopf bifurcations}\label{subsec:SuperSubHopf}

In order to determine whether a Hopf bifurcation is super- or subcritical, the
eigenvalues and eigenvectors about the bifurcation point must first be found.
Among the possible reasons for seeking one or the other kind are that
supercritical Hopf bifurcations can be used for resonant nonlinear
amplification of small periodic signals \cite{Wies86}, and subcritical Hopf
bifurcations are likely to indicate the presence of limit cycles that coexist
with other attractors. The latter type of scenario may give rise to observable
`quantum jumps' among non-fixed point attractors, which would be an interesting
generalization of the predictions of Refs.~\cite{Alsi91,Mabu98}. We thus
believe that the theory for distinguishing super- and subcritical Hopf
bifurcations merits an extended discussion. Note that our expressions below and
in the Appendix correct some apparent misprints in Ref.~\cite{GangNingHaken90},
with minor changes of notation.

At a Hopf bifurcation, the linearized system (\ref{eq:MBJ}) has a pair of pure
imaginary eigenvalues $\lambda_{1,2} = \pm i\omega$, with the frequency
$\omega$ determined by
\begin{equation}\label{eq:HopfFreq}
\omega^2 = \frac{a_1 a_4 - a_5}{a_1 a_2 - a_3}.
\end{equation}
Thus, the characteristic equation (\ref{eq:characteristic}) can be factored as
\begin{equation}\label{eq:characteristicHopf}
(\lambda^2 + \omega^2) (\lambda^3 + a_1 \lambda^2 + b_2 \lambda + b_3) = 0
\end{equation}
with
\begin{eqnarray}
b_2 = a_4 / \omega^2,\quad \mathrm{and} \quad
b_3 = a_5/\omega^2.
\end{eqnarray}
Solving for the other three eigenvalues yields
\begin{equation}\begin{array}{ccc}\label{eq:hopfevalues}
\lambda_3 &=& \nu w - p/(3 \nu w) - a_1/3\\
\lambda_4 &=& \nu^* w - p/(3 \nu^* w) - a_1/3\\
\lambda_5 &=& w - p/(3 w) - a_1/3
\end{array}
\end{equation}
where the variables
\begin{equation}
\begin{array}{ccc}
\nu &=& (-1 + i\sqrt{3} ) / 2 \\
w &=& \left\{q/2 + \sqrt{(q/2)^2 + (p/3)^3}\right\}^{1/3}\\
q &=& -2 a_1^3 /27+ a_1 b_2 / 3 - b_3\\
p &=& -a_1^2 / 3 + b_2
\end{array}
\end{equation}
are determined from the solution to the cubic equation embedded in
(\ref{eq:characteristicHopf}). Following the approach in \cite{GangNingHaken90}
(and noting several corrections), the system eigenvectors, $\mathbf{\alpha_i}$,
may be found in terms of the $\lambda_i$ by solving the linearized dynamics
\begin{equation}
\mathbf{J} \mathbf{\alpha_i} = \lambda_i \mathbf{\alpha_i}
\end{equation}
where $\mathbf{J}$ is the Jacobian in (\ref{eq:MBJ}). Expressing the results in
terms of the $\lambda_i$, one arrives at
\begin{widetext}
\begin{equation}\label{eq:evectors}
\mathbf{\alpha_i} = \left(
\begin{array}{c}
\exp{(i\phi_i)}\\
\exp{(-i\phi_i)}\\
-\exp{(i\phi_i)}\left[k(1+i\Theta)+\lambda_i \right]/(2Ck)\\
-\exp{(-i\phi_i)}\left[k(1-i\Theta)+\lambda_i \right]/(2Ck)\\
-\exp{(i\phi_i)}\left\{2CD_{ss}k+\left(1+i\Delta
+\lambda_i\right)\left[k(1+i\Theta)
+ \lambda_i\right]\right\}/(2Ckx)
\end{array}
\right)
\end{equation}
\end{widetext}
where the phase factor, $\exp{(i\phi_i)}$, is chosen to preserve symmetry in
the components and is given by
\begin{equation}
e^{i\phi_i} = \sqrt{\frac{x\left\{2CD_{ss}k+(1-i\Delta +\lambda_i)[k(1-i\Theta)
+\lambda_i]\right\}}{x^*\left\{2CD_{ss}k+(1+i\Delta +\lambda_i)[k(1+i\Theta)
+\lambda_i]\right\}}}.
\end{equation}
\par
The set of eigenvectors, $\{\mathbf{\alpha_i}\}$, define a linear
transformation of variables such that dynamical equations about steady-state
contain no linear cross couplings. The old variables are related to the new
variables through the relation
\begin{equation}\label{eq:Diagonalization}
\mathbf{\delta q} \, = \widetilde{\alpha}  \, \mathbf{z}
\end{equation}
where $\widetilde{\alpha} = \left[\mathbf{\alpha_1\;\alpha_2 \;\alpha_3
\;\alpha_4 \;\alpha_5}\right]$ is the transformation matrix comprised of the
system eigenvectors, and  $\mathbf{\delta q} = \{\delta x,\;\delta x^*,\;\delta
p,\;\delta p^*,\;\delta D\}^T$ are the fluctuations about steady-state.
Starting from the dynamical equations for the fluctuations about steady-state
\begin{eqnarray}\label{eq:MBfluctuations}
\dot{\delta x} &=& -k \left[(1+i\Theta) \delta x  + 2 C \delta p\right]\nonumber \\
\dot{\delta x^*} &=& (\dot{\delta x})^*\nonumber\\
\dot{\delta p} &=& - (1+i\Delta) \delta p + \delta x \delta D  \\
&& + \ \left\{\delta x D_{ss} + x_{ss} \delta D\right\} \nonumber\\
\dot{\delta p^*} &=& (\dot{\delta p})^*\nonumber\\
\dot{\delta D} &=& -\gamma \left[\delta D + (\delta x^*\delta p+\delta p^*\delta x)/2 \right]
\nonumber \\
&& + \ \left\{\gamma/2 (\delta x^* p_{ss} + x_{ss}^* \delta p + p_{ss}^* \delta
x + \delta p^* x_{ss})\right\} , \nonumber
\end{eqnarray}
the transformation of coordinates $\mathbf{z} = \widetilde{\beta}
\,\mathbf{\delta q}$, with $\widetilde{\beta} = \widetilde{\alpha}^{-1}$,
eliminates any linear coupling between variables
\begin{eqnarray}
\dot{z}_j &=& \lambda_j z_j + \beta_{j3} \delta x\delta D + \beta_{j4} \delta x^*
\delta D \nonumber\\
&& \ -\ \gamma \beta_{j5} (\delta x^* \delta p + \delta x\delta p^*)/2 \ .
\end{eqnarray}
Finally, after utilizing the transformation (\ref{eq:Diagonalization}), this
equation may be expressed in terms of the ``diagonalized" coordinates alone:
\begin{eqnarray}\label{eq:MBDiag}
\dot{z}_j &=& \lambda_j z_j + \sum_{k,l = 1}^5 \left[\beta_{j3} \alpha_{1k}
\alpha_{5l}+ \beta_{j4}\alpha_{2k}\alpha_{5l} \right.\nonumber \\
&& \left. - \ \gamma \beta_{j5} (\alpha_{2k}\alpha_{3l} + \alpha_{1k}
\alpha_{4l})/2\right] z_k z_l \ .
\end{eqnarray}
\par
Having converted the system into the form (\ref{eq:MBDiag}), the dynamics about
a Hopf bifurcation may be reduced onto a center manifold \cite{Guckenheimer}:
since the system dynamics will be dominated by the ``slow" variables, $z_1$ and
$z_2$, the flow of the differential equation may be locally approximated on the
surface generated by $z_1$ and $z_2$, with the ``fast" variables,
$z_{j=3,4,5}$, represented by a local graph $z_j = A_j(z_1,z_2)$. Furthermore,
the local graph, $A_j(z_1,z_2)$, may be approximated by a power series
expansion
\begin{equation}\label{eq:CenterMan}
z_j = a_{20}(j)z_1^2+a_{11}(j)z_1z_2+a_{02}(j)z_2^2+...\, ,\;j = 3,4,5 \; ,
\end{equation}
so that the reduced dynamics may be approximated by
\begin{equation}\label{eq:CenterManDynamics}
\dot{z}_j \approx 2 i \omega a_{20}(j)z_1^2-2 i \omega a_{02}(j)z_2^2+...\,
,\;j = 3,4,5 \, .
\end{equation}
The coefficients in (\ref{eq:CenterMan}) are determined by substituting
equations (\ref{eq:CenterMan}) and (\ref{eq:CenterManDynamics}) into the exact
dynamics (\ref{eq:MBDiag}) and equating like powers in $z_1^n z_2^m$.
\par
With aid of the local graph (\ref{eq:CenterMan}), the dynamics may be reduced
onto the center coordinates associated with eigenvalues having zero real part:
\begin{equation}\label{eq:ReducedDynamics}
\dot{z}_1 = i\omega z_1 + b_{20}(1) z_1^2 + b_{11}(1) z_1 z_2 + b_{02}(1) z_2^2
+ b_{21}(1) z_1^2 z_2 + ...
\end{equation}
with $z_2 = z_1^*$ near the Hopf bifurcation. Writing $z_1 = u + i v$, with
$\{u,v\} \in \mathbb{R}$, the reduced dynamics of the complex variable $z_1$
may be expressed by a set of real coupled equations
\begin{equation}\label{eq:ReducedReal}
\begin{array}{lll}
\dot{u}&=& -\omega v + F(u,v)\\
\dot{v}&=& \omega u + G(u,v)
\end{array}\ ,
\end{equation}
with $F$ and $G$ comprised of terms nonlinear in $u$ and $v$. It can be shown
\cite{Guckenheimer}, that there exists a smooth change of variables which will
put equation (\ref{eq:ReducedReal}) into the normal form
\begin{equation}\label{eq:ReducedPolar}
\begin{array}{lll}
\dot{r} &=& \eta_3 r^3 + \eta_5 r^5 + ...\\
\dot{\theta} &=& \omega + \epsilon_2 r^2 + \epsilon_4 r^4 + ...
\end{array}\ ,
\end{equation}
with the stability of the bifurcation governed by the sign of $\eta_3$. For
$\eta_3 <0$ the bifurcation is supercritical, and a stable, small amplitude
limit cycle  is born about the newly destabilized steady-state. For $\eta_3
>0$, the bifurcation is subcritical, and no such small amplitude cycle is
created. In this case, the system may be thrown far away from the steady-state
solution, onto either a limit cycle, a different branch of the steady-state
curve, or some other attractor.

Remaining details of the calculation of $\eta_3$ are relegated to the Appendix.

\section{Quantum signatures of bistability and limit cycles}

In this section we briefly review some computational tools that can be used to
search for evidence of the semiclassical attractors in the quantum model. In
the following section we present numerical results in which these tools are
applied at several interesting points in the semiclassical bifurcation set.

\subsection{Steady-state Q-function}
The Q-function (for the intracavity optical field) \cite{Wall95} can be used to
characterize the steady-state behavior of the driven atom-cavity system. The
benefits of using the Q-function, over the other common quasi-probability
distributions, are twofold. Firstly, the Q-function is positive-semidefinite,
and is therefore better suited for making comparisons to classical probability
distributions (such as the stationary distribution of a classical model with
noise). Secondly, the value of the Q-function has a very simple interpretation
in terms of coherent states $|\alpha\rangle$ of the intracavity field, since
for any given density operator $\rho$ we have $Q(\alpha) = \langle \alpha |
\rho|\alpha \rangle$. Thus $Q(\alpha)$ may strictly be thought of as a
probability, which further lends to the utility of the Q-function as a tool for
making comparisons between the quantum and semiclassical descriptions of the
equilibrium behavior.

In practice we can find the steady-state density operator $\rho_{ss}$ of the
master equation (\ref{eq:JCMaster}) simply by setting the terms on its
right-hand side to zero and solving the resulting algebraic equation. We can
then use $Q_{ss}(\alpha) = \langle \alpha | \rho_{ss}|\alpha \rangle$, where
there is an implied partial trace over the atomic degrees of freedom.

One naively expects that the Q-function should be bimodal when atom-cavity
parameters are chosen in a bistable region of the Maxwell-Bloch equations
\cite{Sava88}. Likewise, the existence of a limit cycle should give rise to a
ring-shaped Q-function. Below we will show examples of such features, but we
will also want a method for visualizing any possible coherent dynamics or large
fluctuations buried within these stationary distributions. For example we want
to be able to show that a ringlike Q-function indicates an intracavity field
amplitude that oscillates coherently, as opposed to randomly diffusing around
the circle. In the case of bistability we would also like to see the switching
timescale and to examine the sharpness of the switching events.

\subsection{Quantum Trajectories}
To visualize dynamics in the quantum model in an experimentally-relevant way,
we turn to the method of quantum trajectory simulations
\cite{CarmichaelOpenSys,Wise93a}.

The master equation (\ref{eq:JCMaster}) generates predictions of the
unconditioned state of the atom-cavity system. That is, the solutions $\rho(t)$
of the master equation represent the knowledge we can have of the evolving
system state without utilizing the information obtainable via real-time
measurements of the output fields (cavity transmission and atomic
fluorescence). Hence we may gain further insight into the dynamics by
considering quantum predictions regarding the photocurrents; fortunately, the
same theory used to derive the master equation provides a powerful set of tools
for statistically-faithful sampling of continuous measurement records
\cite{CarmichaelOpenSys,Wise93a}, and tells us how to interpret them as
real-time observations of the intracavity dynamics \cite{Bout06}. Here we will
use such `quantum trajectory' methods for monte carlo simulations of the
photocurrent generated by homodyne detection of the cavity output field
\cite{Wise93a}.

For the case of homodyne detection of the output fields, the stochastic
Schr\"odinger equation (SSE) governing the evolution of the (unnormalized)
conditional state vector, $|\psi_c\rangle$, is given by %
\be\label{eq:SSE}
\begin{array}{lclclc}
d |\psi_c\rangle &=& -i\left(\mathcal{H} - i \kappa a^\dag a - i\gammaperp
\sigma_+ \sigma_-\right)|\psi_c\rangle + \\
&& \ \sqrt{2\kappa} a e^{i\phi_1} |\psi_c\rangle d Q_1 + \\
&& \ \sqrt{2\gammaperp} \sigma_- e^{i\phi_2} |\psi_c\rangle d Q_2 \ ,
\end{array}
\ee%
with the Hamiltonian, $\mathcal{H}$, given in Eq.~(\ref{eq:JCHamiltonian}), and the
phase factor $e^{i\phi_{1,2}}$ determining the field quadrature being measured
(\textit{e.g.} $\phi_{1,2} = 0$ for measurement of the amplitude quadrature and
$\phi_{1,2} = -\pi/2$ for the phase quadrature.) The measured homodyne photocurrents,
$I_{hom_{1,2}} = \frac{dQ_{1,2}}{dt}$, which are, respectively, the homodyne
photocurrents associated with the cavity and atomic decay channels, are calculated
using%
\bea\label{eq:Photocurrent}
dQ_1 = \sqrt{2\kappa}\frac{\langle a^\dag  e^{-i\phi_1} + a
e^{i\phi_1}\rangle_c}{\langle \psi_c |\psi_c\rangle} dt + dW_1 \ ,\\
dQ_2 = \sqrt{2\gammaperp}\frac{\langle \sigma_+ e^{-i\phi_2} + \sigma_-
e^{i\phi_2}\rangle_c}{\langle \psi_c |\psi_c\rangle} dt + dW_2 \ ,
\eea%
where $dW_{1,2}$ are independent Wiener increments satisfying $\langle dW
\rangle = 0$ and $\langle (dW)^2 \rangle = dt$. Numerical integration of
equations (\ref{eq:SSE},\ref{eq:Photocurrent}) are performed using the
stochastic integration routine incorporated in the Quantum Optics Toolbox
\cite{Tan99} for MATLAB.

In any practical experiment, full measurement of the atomic spontaneous
emission is not actually feasible as this would require a detector covering
nearly $4\pi$ steradians of solid angle. Fortunately, the cavity-output
homodyne photocurrent $I_{hom_1}$ generated by monte carlo integration of the
above SSE (considered on its own without any reference to the corresponding
$I_{hom_2}$) is sampled from the same law as the photocurrent one would see in
an experiment in which the atomic decay channel was not measured at all
\cite{vanH05}. We therefore make use of such photocurrent simulations below in
our discussion of single-atom bistability and Hopf bifurcations. One should
appreciate however that the conditional state $\vert\psi_c\rangle$ propagated
by the SSE is then merely an internal variable of the monte carlo simulation,
and not something that could actually be reconstructed (in a recursive
estimation sense) from just the cavity-output photocurrent. The best one could
do along the latter lines, without assuming high-efficiency observation of the
cavity decay channel, would be to utilize the corresponding stochastic master
equation (SME) \cite{Wise93a} as an optimal quantum filter \cite{Bout06}. Even
in a purely theoretical discussion one would like to utilize the SME (as in
Ref.~\cite{Mabu98}) to generate not only realistic photocurrent samples, but
also the conditional quantum states that one could in principle generate from
them via recursive filtering. Unfortunately such numerical procedures are very
computationally intensive. As we find that adequate indications of the dynamics
underlying bimodal and ringlike Q-functions are provided by the photocurrents
alone, we have limited our efforts to SSE-based simulations.

\section{Numerical results}

\subsection{Absorptive bistability}
\begin{figure}[tb!]
\centering
{\includegraphics[width=0.475\textwidth]{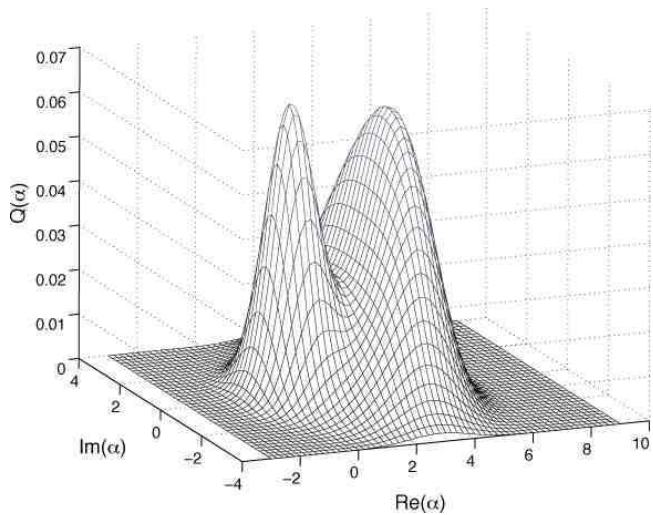}}
\caption{\label{fig:AbsorptiveBistabilityQfuncOnly} Steady-state $Q(\alpha)$
for the parameter values in Fig.~\ref{fig:AbsorptiveBistabilityMBSim}, and
drive field $y=11.3$.}
\end{figure}

In Fig.~\ref{fig:AbsorptiveBistabilityMBSim} we plot the steady-state
intra-cavity field magnitude vs.\ drive field predicted by the (dimensionless)
semi-classical equations (\ref{eq:MBEOS}) for the case of purely absorptive
bistability ($\Theta=\Delta=0$). These parameter values correspond to
$g=1.41,\kappa=.1,\gammapar=2,\Delta_a=0,\Delta_c=0$, in the master equation
(\ref{eq:JCMaster}), and a saturation photon number, $n_{0}=0.25$. In
Fig.~\ref{fig:AbsorptiveBistabilityQfuncOnly} we plot the bimodal Q-function
obtained from the steady solution to the master equation for a drive field
$y=11.3$ ($\mathcal{E}=0.565$), chosen such that the integrated probabilities in
each mode of the Q-function are approximately equal. While this Q-function
indicates that the quantum dynamics show bistable behavior, it is interesting
to note that the master equation produces this bimodal distribution for a drive
field where the mean-field equations do not predict bistability (the lower
branch in Fig.~\ref{fig:AbsorptiveBistabilityMBSim} disappears at
$y\approx11.1$). In fact, the Q-function distributions over most of the
semi-classically bistable region are not bimodal, and only become so for
$11\lesssim y\lesssim 11.5$.

\begin{figure}[bth!]
\centering
{\includegraphics[width=0.475\textwidth]{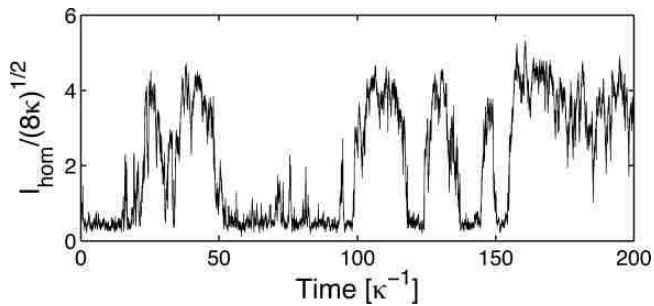}}
\caption{\label{fig:AbsorptiveBistabilityTrajectory_113} Homodyne photocurrent
from quantum trajectory simulation for parameter values in
Fig.~\ref{fig:AbsorptiveBistabilityQfuncOnly}.}
\end{figure}

In Fig.~\ref{fig:AbsorptiveBistabilityTrajectory_113} we plot the photocurrent
(\ref{eq:Photocurrent}) corresponding to a measurement of the amplitude
quadrature of the cavity output field. As expected, the field localization
brought about by the continuous homodyne measurement causes the signal amplitude to
switch, at stochastic intervals, between values consistent with the two peaks
in $Q(\alpha)$. 

\subsection{Supercritical Hopf bifurcation}
\begin{figure}
\centering 
{\includegraphics[width=0.475\textwidth]{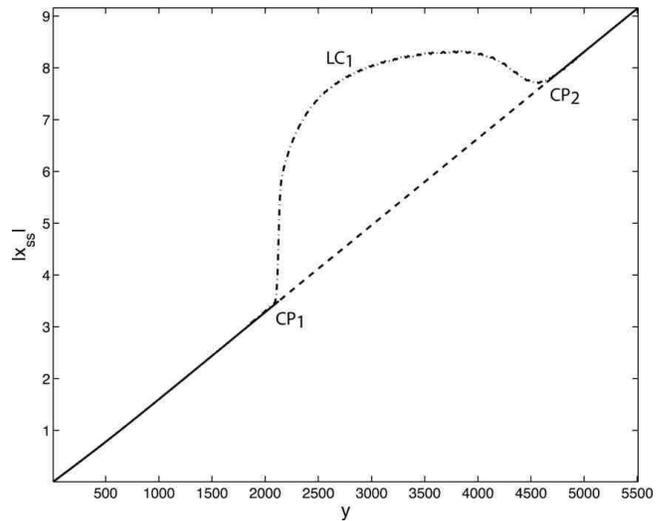}}
\caption{\label{fig:RingModeMBSim} Semiclassical calculation of the intracavity
steady-state field magnitude $|x_{ss}|$ versus drive field $y$. The parameter
values are: $C=50,k=.01,\gamma=2,\Theta=-600$, and $\Delta=1.25.$ The upper
dashed-dotted curve ($LC_1$)represents the steady-state oscillation maximum
(steady-state magnitude plus mean value) associated with the limit cycle formed
when the fixed point becomes unstable due to a Hopf bifurcation (at $CP_1$ and
$CP_2$). The lower dashed curve is unstable.}
\end{figure}

In Fig.~\ref{fig:RingModeMBSim} we plot the equilibrium attractors of the
mean-field dynamics (\ref{eq:MBdimless}) for a case where the steady-state
fixed points predicted in (\ref{eq:MBEOS}) undergo supercritical Hopf
bifurcations. Starting on the lower (upper) branch of the steady-state curve,
as the drive field $y$ is swept through the critical point $CP_1$ ($CP_2$) the
fixed point is destabilized by a small amplitude limit cycle, $LC_1$, which
grows in amplitude, peaking at $y\approx2800$, until finally recombining with
and restabilizing the fixed point at $CP_2$ ($CP_1$). To represent the
oscillatory solution that is born out of the bifurcation, we plot the
steady-state \textit{maximum} field magnitude for a state localized on the
stable limit cycle, and denote this as $LC_1$. Thus, the plotted curve $LC_1$
essentially represents the amplitude \textit{plus} mean value of the limit
cycle.
\par
\begin{figure}[tb!]
\centering 
{\includegraphics[width=0.475\textwidth]{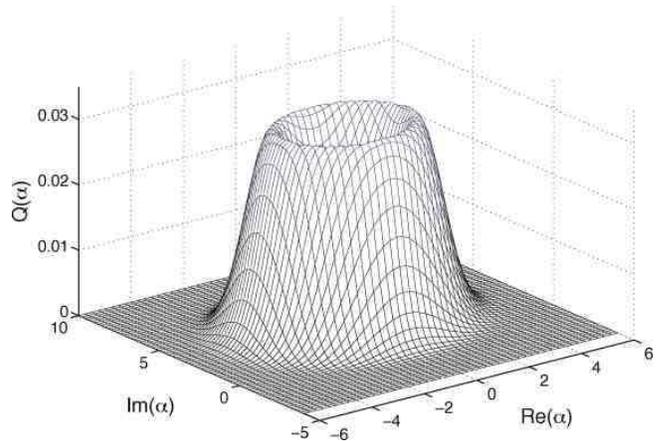}}
\caption{\label{fig:RingModeQfuncOnly} Steady-state $Q(\alpha)$ for the
parameter values in Fig.~\ref{fig:RingModeMBSim}, and drive field $y=2800$.}
\end{figure}
\begin{figure}[bt!]
\centering 
{\includegraphics[width=0.475\textwidth]{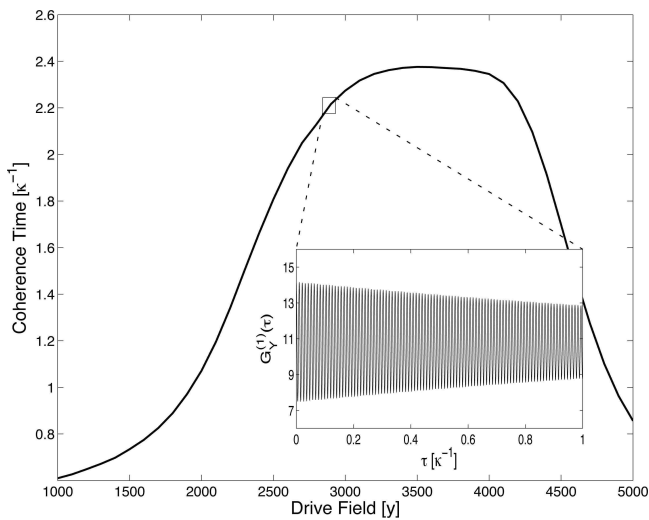}}
\caption{\label{fig:RingModeCoherenceCorrelation} Coherence times calculated
from steady-state autocorrelation function, $G_Y^{(1)}(\tau)$,  of the quantum
mechanical phase quadrature.  Parameter values same as in
Fig.~\ref{fig:RingModeMBSim}. The coherence time is estimated by fitting a
damped sinusoid to $G_Y^{(1)}(\tau)$. Inset: simulated $G_Y^{(1)}(\tau)$ for
the drive field in Fig.~\ref{fig:RingModeQfuncOnly}.}
\end{figure}

The parameter values used in Fig.~\ref{fig:RingModeMBSim} correspond to
$g=1,\kappa=.01,\gammapar=2,\Delta_a=1.25,\Delta_c=-6$ in
Eq.~(\ref{eq:JCMaster}), and a saturation photon number $n_{0}=0.5$. Using
these values, we compute $Q(\alpha)$ for $y=2800$ ($\mathcal{E} = 19.8$), where
the limit cycle amplitude is maximal. The result is plotted in
Fig.~\ref{fig:RingModeQfuncOnly}. The ring-like shape of the distribution is
consistent with oscillation of a coherent state in the intracavity field. This
interpretation is further supported by the inset in
Fig.~\ref{fig:RingModeCoherenceCorrelation}, where we plot the autocorrelation
function $G_Y^{(1)}(\tau) = \langle Y(\tau) Y(0) \rangle-\langle
Y(0)\rangle^2$, computed using the quantum regression theorem \cite{Wall95},
where $Y = \frac{i}{2}(a^\dag-a)$ is the phase quadrature amplitude operator of
the intracavity field. In addition, Fig.~\ref{fig:RingModeCoherenceCorrelation}
displays the coherence time of the steady-state quantum oscillations over a
range of drive fields. The results indicate that the coherence times depend
strongly on the amplitude of the limit cycle, $LC_1$, which is again consistent
with the idea of an oscillating coherent state for the intracavity field.

\par
\begin{figure}[tb!]
\centering 
{\includegraphics[width=0.475\textwidth]{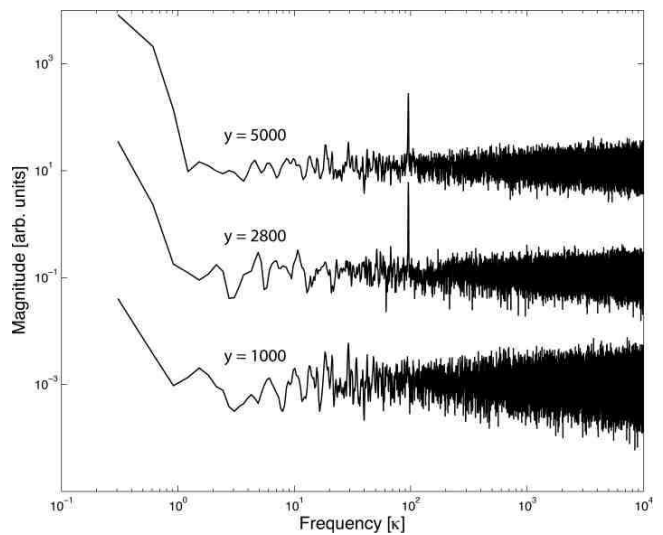}}
\caption{\label{fig:Spectra_1000_2800_5000} Power spectra computed from
simulated photocurrents for amplitude-quadrature homodyne detection of the
cavity output field, using parameter values as in Fig.~\ref{fig:RingModeMBSim}.
Lowest plot: $y = 1000$, below the bifurcation point; middle plot: $y=2800$,
where the classical oscillation amplitude is maximal; upper plot: $y=5000$,
where the semiclassical model no longer predicts a limit cycle.}
\end{figure}

It can be seen clearly from the inset of
Fig.~\ref{fig:RingModeCoherenceCorrelation} that the limit cycle comprises an
oscillation of the intracavity field at a frequency much higher than $\kappa$.
It should thus be difficult to see the oscillation directly in the broadband
photocurrent generated by amplitude-quadrature homodyne detection of the cavity
output field. In Fig.~\ref{fig:Spectra_1000_2800_5000} however we plot several
power spectra of photocurrent records generated in quantum trajectory
simulations. For $y=1000$ (below $CP_1$ in Fig.~\ref{fig:RingModeMBSim}) the
spectrum shows little or no sign of a coherent peak, but for $y=2800$ we see
that homodyne detection of the field amplitude reveals clear evidence of the
limit-cycle oscillation. This demonstrates at least a basic correspondence with
the semiclassical predictions shown in Fig.~\ref{fig:RingModeMBSim}. At
$y=5000$ (above $CP_2$ in Fig.~\ref{fig:RingModeMBSim}), however, we see that
the quantum model still exhibits strong oscillations even though the
semiclassical model predicts a fixed point solution. This persistence of the
oscillatory behavior at both higher and lower driving fields can also be seen
in Fig.~\ref{fig:RingModeCoherenceCorrelation}.

\subsection{Subcritical Hopf bifurcation}
\begin{figure}[tb!]
\centering 
{\includegraphics[width=0.475\textwidth]{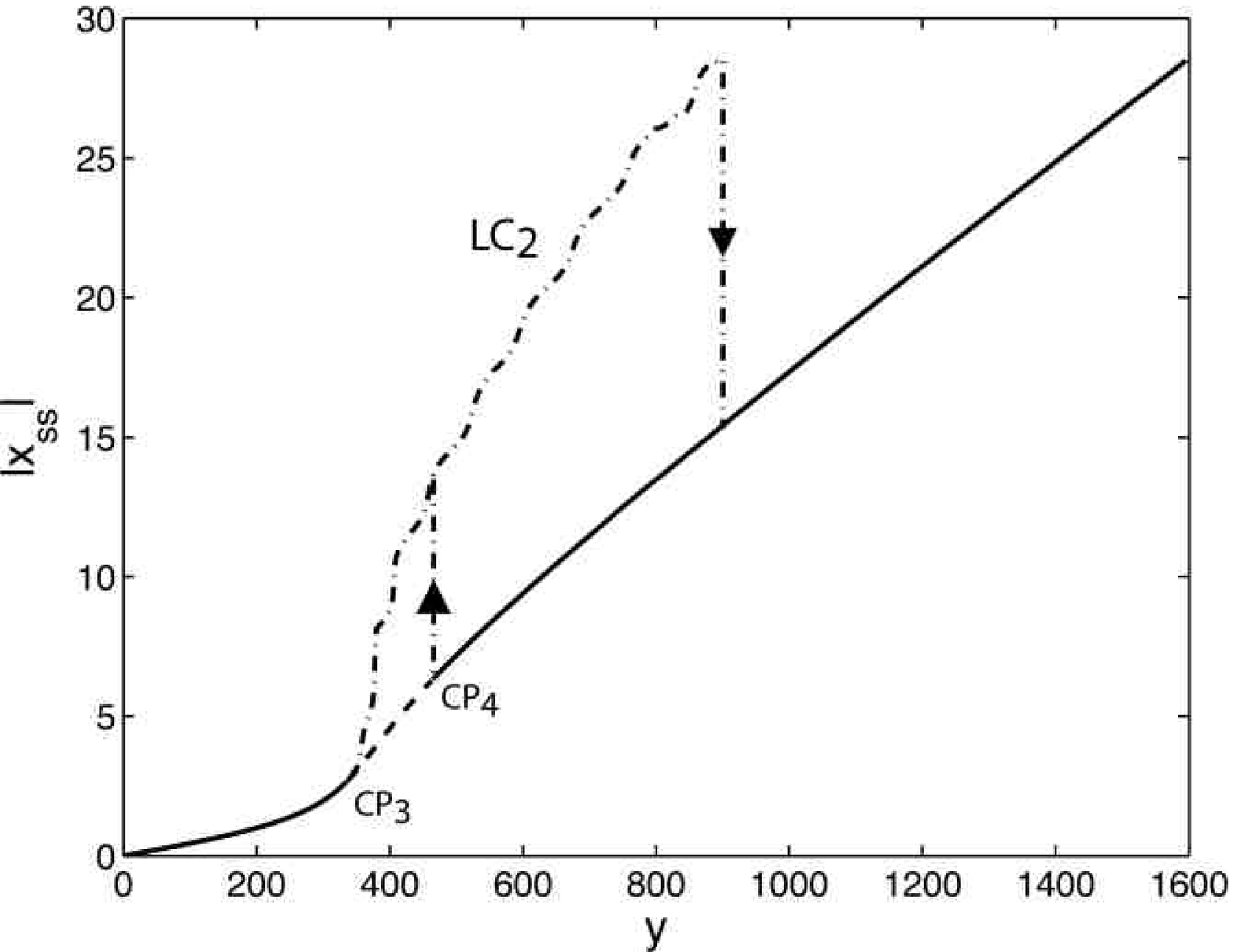}}
\caption{\label{fig:SwirlyModeMBSim} Semiclassical calculation of the
intracavity steady-state field magnitude $|x_{ss}|$ versus drive field $y$. The
parameter values are: $C=200,k=.05,\gamma=2,\Theta=-55,$ and $\Delta=2$. The
upper dashed-dotted curve ($LC_2$) represents the steady-state oscillation
maximum (steady-state magnitude plus mean value) associated with the limit
cycle formed when the fixed point becomes unstable due to a Hopf bifurcation.
The critical point $CP_3$ is supercritical, whereas the point $CP_4$ is
subcritical. Note the coexistence of a stable fixed point and limit cycle in
the range $475\lesssim y \lesssim925$. The lower dashed curve is unstable. }
\end{figure}

In Fig.~\ref{fig:SwirlyModeMBSim} we plot the steady-state solutions for a
parameter regime where the mean-field equations predict a subcritical Hopf
bifurcation. The solid (dashed) curve corresponds to the stable (unstable)
fixed points predicted by Eq.~(\ref{eq:MBEOS}), whereas the attractor $LC_2$
(plotted dashed-dotted) corresponds to a stable limit cycle. Beginning on the
upper stable branch of fixed points, as the drive field is swept through the
critical point $CP_4$, the system undergoes a subcritical Hopf bifurcation. In
the range $475\lesssim y \lesssim925$, the semiclassical equations predict
coexistence of a stable fixed point and limit cycle, which is a common
signature of subcritical bifurcations. Note that at $y\approx925$ the limit
cycle $LC_2$ is destabilized but the fixed point is not. The two arrowed lines
in Fig.~\ref{fig:SwirlyModeMBSim} do not represent solutions to the equations,
but simply indicate which attractor a destabilized state will seek.
\par
\begin{figure}[tb!]
\centering 
{\includegraphics[width=0.475\textwidth]{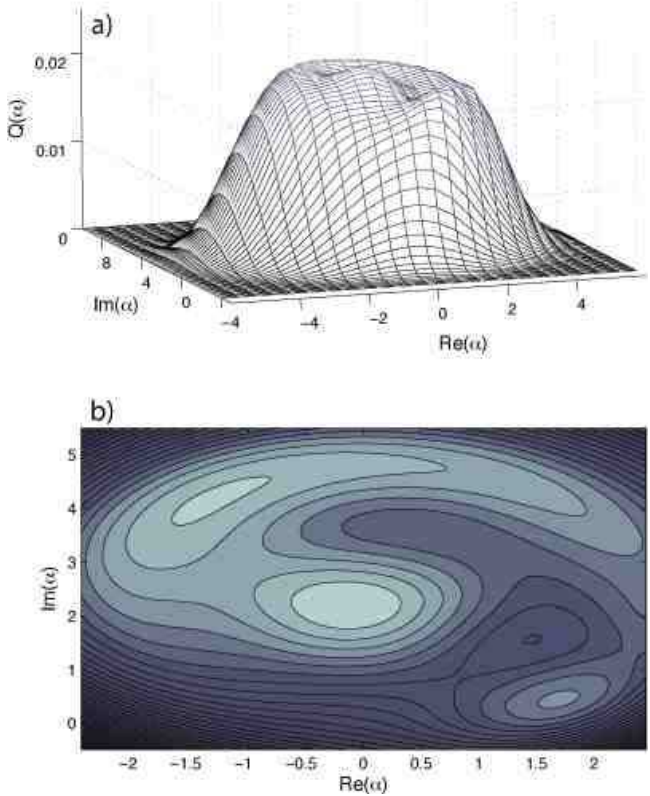}}
\caption{\label{fig:SwirlyModeQFuncFaceted} Steady-state surface (a) and
contour plots (b) of $Q(\alpha)$ for the parameter values in
Fig.~\ref{fig:SwirlyModeMBSim}, and drive field $y=1000$.}
\end{figure}

The parameter values used in Fig.~\ref{fig:SwirlyModeMBSim} correspond to
$g=4.47,\kappa=.05,\gammapar=2,\Delta_a=2,\Delta_c=-2.75$, and a saturation
photon number $n_{0}=0.025$. The small size of $n_0$ indicates that the
qualitative behavior of steady-state solutions to the master equation will be
dominated by quantum fluctuations (over the dynamics implied by the mean field
equations.) We see that this interpretation is justified by the plot of
$Q(\alpha)$ in Fig.~\ref{fig:SwirlyModeQFuncFaceted}, at a drive field
$y=1000$, near the amplitude maximum of $LC_2$. The features in the surface
plot (Fig.~\ref{fig:SwirlyModeQFuncFaceted}a) are hardly profound. The contour
plot in Fig.~\ref{fig:SwirlyModeQFuncFaceted}b is somewhat more elucidating,
and shows signs of the coexistence of oscillatory, albeit asymmetric, and fixed
coherent states. These `blurry' results are not surprising, as the small $n_0$,
and relation (\ref{eq:FieldScaling}), imply that, in the quantum case, the
structure implied by the limit cycle and fixed points in
Fig.~\ref{fig:SwirlyModeMBSim} should be heavily affected by fluctuations.
\par
\begin{figure}
\centering
{\includegraphics[width=0.475\textwidth]{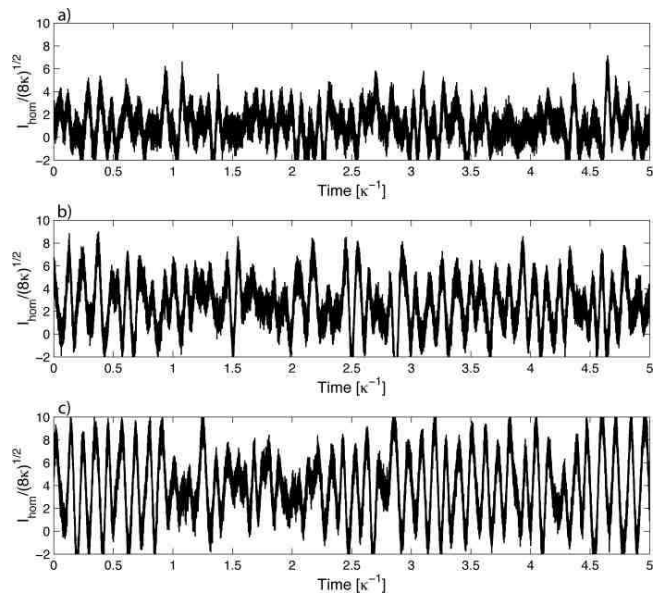}}
\caption{\label{fig:SwirlyModeTrajectories_500_1000_1400} Homodyne photocurrent
from quantum trajectory simulations with parameter values in
Fig.~\ref{fig:SwirlyModeMBSim}. (a) drive field $y = 500$, near the subcritical
bifurcation point; (b) for drive field $y=1000$, near the oscillation
amplitude maximum; (c) for drive field $y=1400$, beyond the region of
semiclassical bistability.}
\end{figure}
Fig.~\ref{fig:SwirlyModeTrajectories_500_1000_1400} suggests, however, that the
field localization provided by continuous homodyne measurement of the phase
quadrature can reveal signatures of bistability in the photocurrent. In
particular, in Fig.~\ref{fig:SwirlyModeTrajectories_500_1000_1400}b we plot a
typical trajectory for a drive field $y=1000$, as in
Fig.~\ref{fig:SwirlyModeQFuncFaceted}. While the contrast is marginal, as it
would have to be given the discussion above, the qualitative appearance of this
simulated photocurrent record is that of oscillations interrupted by brief
periods of stationary noise (which one could attribute to transient
localization on the fixed point). Such intermittency can also be seen in
Fig.~\ref{fig:SwirlyModeTrajectories_500_1000_1400}c, which corresponds to a
drive field, $y=1400$, well past the semiclassical region of multi-stability.

\section{Conclusions}

In this paper we have considered bifurcation phenomena as focal points for the
investigation of quantum-(semi)classical correspondence in cavity nonlinear
optics. We presented a general approach to the characterization of interesting
points in the semiclassical (Maxwell-Bloch) bifurcation set, starting from the
formal methods of Refs.~\cite{GangNingHaken90,GNH90b} (but with corrections to
some of their equations as printed) and incorporating numerical simulations of
the cavity QED master equation and (homodyne) stochastic Schr\"{o}dinger
equation. This approach leads to the prediction of self-oscillation and
bifurcation-type behavior in an experimentally accessible parameter regime for
single-atom cavity QED, under driving conditions in which the mean intracavity
photon number is of order ten. It is interesting to see that such a wide range
of input-output characteristics are supported in such a small state space.

The results of our numerical simulations point to a number of questions for
further study. For example one would like to understand, in physical terms,
what determines the correlation timescales found in
Fig.~\ref{fig:AbsorptiveBistabilityTrajectory_113} (bistability) and
Fig.~\ref{fig:RingModeCoherenceCorrelation} (stable limit cycle). In scenarios
where switching occurs between output-signal characteristics associated with
different semiclassical attractors (as in
Fig.~\ref{fig:AbsorptiveBistabilityTrajectory_113} and
Fig.~\ref{fig:SwirlyModeTrajectories_500_1000_1400}), one would also like to
know what determines the attractors' relative stability and how the switching
events are initiated (in the sense of
Refs.~\cite{Alsi91,Mabu98,Well01,vanH05}). Can we understand the physical
dynamics that give rise to limit-cycles in single-atom cavity QED (as in
Ref.~\cite{Alsi91}), and why they are destabilized in subcritical Hopf
bifurcations? Such investigations could be taken as a starting point for the
development of quantum feedback control strategies \cite{Dohe00,vanHIEEE} to
stabilize desired fixed points \cite{Mack93,Yaow99} or limit cycles, or for
inducing switching between them with minimum time or dissipated energy (as
would be required for the types of optical signal processing applications
mentioned in the introduction). More generally, one can ask about the
applicability of ideas from classical bifurcation control \cite{Abed86,Chen00}.
In this context, single-atom cavity QED provides an interesting model system
for investigations of quantum feedback control far beyond the linear-gaussian
regime \cite{Wise05,Stoc04a}.

Returning finally to the issue of correspondence, one curious detail of our
simulation results is that near hysteresis loops, bimodal behavior in the
quantum model is often seen to persist well above (at higher driving fields
than) the upper switching points of the Maxwell-Bloch equations, where
saddle-node or subcritical bifurcations occur. We have also seen a case
(supercritical Hopf) in which oscillatory behavior occurs over a wider
parameter range than is predicted by the semiclassical model. It is certainly
possible that these effects may be explainable by analogy with noise-induced
`postponement' \cite{Robi85,Fron87}, `advancement' or `precursors'
\cite{Wies85} in classical nonlinear systems, or with some other known effect
in classical random dynamics \cite{Hors84,Arno92,Arno99,Wack99}. If this is the
case, it will be interesting to see what clues such analogies provide towards
the development of physically-motivated stochastic extensions of the
Maxwell-Bloch equations that can capture global (in the dynamical systems
sense) effects of quantum fluctuations (thus going beyond what is possible with
local-linearization approaches, as in Ref.~\cite{CarmichaelNoAdiabatic}). In
any case it will be natural to try to relate the machinery of `P-bifurcation'
analysis \cite{Arno98,Zeem88} to our steady-state Q-functions. Of course if
sensible analogies with postponement in classical noise-driven systems cannot
be established, it will be tempting to ask whether coherence or atom-field
entanglement play any significant role in these or any other cavity QED
bifurcation phenomena.

In any case, we certainly expect that further study of bifurcation phenomena in
single-atom cavity QED will improve our understanding of the way that quantum
fluctuations `enrich' the mean-field phase portrait of coherent nonlinear
dynamical systems.

\begin{acknowledgments}

The authors thank R.~van~Handel for insightful discussions. This work was
supported by the NSF under grant number PHY-0354964.

\end{acknowledgments}

\vbox{\medskip}

\appendix
\section{Calculation of $\eta_3$}

The critical parameter $\eta_3$ can be expressed explicitly
\cite{Guckenheimer,Ning88} in terms of the coefficients in
(\ref{eq:ReducedDynamics}) and the bifurcation frequency $\omega$
\begin{equation}
\eta_3 = \mathrm{Re}[b_{21}(1)] - \frac{1}{\omega} \mathrm{Im} [b_{20}(1)
b_{11}(1)]\; .
\end{equation}
As the calculations for the relevant coefficients are rather tedious and time
consuming, they are included here for posterity:
\begin{eqnarray}\label{eq:aCoeffs}
\begin{array}{lll}
a_{20}(j) &=& b_{20}(j)/(2 i \omega -\lambda_j)\; ,\\
a_{11}(j) &=& - b_{11}(j)/\lambda_j\;,
\end{array} \bigg\}\; j = 3,4,5
\end{eqnarray}
where
\begin{eqnarray}\label{eq:bCoeffs}
b_{20}(j) &=& \beta_{j3} \alpha_{11}\alpha_{51} +
\beta_{j4}\alpha_{21}\alpha_{51}
\nonumber\\
&& - \ \gamma \beta_{j5} (\alpha_{21}\alpha_{31} + \alpha_{11}\alpha_{41})/2\ , \\
b_{11}(j) &=& \beta_{j3}
\left(\alpha_{11}\alpha_{52}+\alpha_{12}\alpha_{51}\right)
\nonumber\\
&& + \ \beta_{j4}\left( \alpha_{21}\alpha_{52}+\alpha_{22}\alpha_{51}\right) \nonumber\\
&& - \ \gamma \beta_{j5} \left[\alpha_{21}\alpha_{32} +
\alpha_{22}\alpha_{31}\right.
\nonumber\\
&& + \ \left. \alpha_{11}\alpha_{42}+\alpha_{12}\alpha_{41}\right]/2 \; .
\end{eqnarray}
Inserting (\ref{eq:CenterMan}), with coefficient $a_{20}(j)$ and $a_{11}(j)$
given by (\ref{eq:aCoeffs}), into the diagonalized equation (\ref{eq:MBDiag})
for $j=1$ yields an expression for the last required coefficient
\begin{widetext}
\begin{equation}\begin{array}{lll}
b_{21}(1)&=& \beta_{13} [ a_{20}(3)\left(\alpha_{12}\alpha_{53}+\alpha_{13}\alpha_{52}\right) + a_{20}(4)\left( \alpha_{12}\alpha_{54}+\alpha_{14}\alpha_{52} \right)+ a_{20}(5) \left( \alpha_{12}\alpha_{55}+\alpha_{15}\alpha_{52} \right) +\\
&& \quad a_{11}(3) \left( \alpha_{11}\alpha_{53} + \alpha_{13}\alpha_{51} \right)+a_{11}(4) \left( \alpha_{11}\alpha_{54}+\alpha_{14}\alpha_{51} \right) + a_{11}(5) \left( \alpha_{11}\alpha_{55}+\alpha_{15}\alpha_{51} \right) ]\\
&+& \beta_{14} [ a_{20}(3)\left(\alpha_{22}\alpha_{53}+\alpha_{23}\alpha_{52}\right) + a_{20}(4)\left(\alpha_{22}\alpha_{54}+\alpha_{24}\alpha_{52}\right)+\\
&& \quad\quad a_{20}(5) \left(\alpha_{22}\alpha_{55}+\alpha_{25}\alpha_{52}\right) + a_{11}(3)\left(\alpha_{21}\alpha_{53}+\alpha_{23}\alpha_{51}\right)+\\
&& \quad\quad\quad a_{11}(4) \left(\alpha_{21}\alpha_{54}+\alpha_{24}\alpha_{51}\right) + a_{11}(5)\left(\alpha_{21}\alpha_{55}+\alpha_{25}\alpha_{51}\right) ]\\
&-&\frac{\gamma}{2} \beta_{15} [ a_{20}(3)\left(\alpha_{22}\alpha_{33}+\alpha_{23}\alpha_{32}+\alpha_{12}\alpha_{43}+\alpha_{13}\alpha_{42}\right)+\\
&& \quad\quad\quad a_{20}(4) \left(\alpha_{22}\alpha_{34}+\alpha_{24}\alpha_{32}+\alpha_{12}\alpha_{44}+\alpha_{14}\alpha_{42}\right)+\\
&& \quad\quad\quad\quad a_{20}(5) \left(\alpha_{22}\alpha_{35}+\alpha_{12}\alpha_{45}+\alpha_{25}\alpha_{32}+\alpha_{15}\alpha_{42}\right)+\\
&& \quad\quad\quad\quad\quad a_{11}(3) \left(\alpha_{21}\alpha_{33}+\alpha_{11}\alpha_{43}+\alpha_{23}\alpha_{31}+\alpha_{13}\alpha_{41}\right)+\\
&& \quad\quad\quad\quad\quad\quad a_{11}(4) \left(\alpha_{21}\alpha_{34}+\alpha_{11}\alpha_{44}+\alpha_{24}\alpha_{31}+\alpha_{14}\alpha_{41}\right)+\\
&& \quad\quad\quad\quad\quad\quad\quad a_{11}(5)
\left(\alpha_{21}\alpha_{35}+\alpha_{11}\alpha_{45}+\alpha_{25}\alpha_{31}+\alpha_{15}\alpha_{41}\right)]
\; .\end{array}
\end{equation}
\end{widetext}

\end{document}